\documentclass[nofootinbib,aps,preprint,longbibliography]{revtex4-1}
\usepackage{amsmath}
\usepackage{bbold}
\usepackage{graphicx}
\usepackage[hidelinks]{hyperref}
\usepackage{bookmark}% faster updated bookmarks
\usepackage{lineno}
\newcommand{\parallelsum}{\mathbin{\!/\mkern-5mu/\!}}

\begin{document}
\title{Geometrical Hall effect and momentum-space Berry curvature from spin-reversed band pairs}

\author{Max Hirschberger$^{1,2,\ddagger}$}
\email{hirschberger@ap.t.u-tokyo.ac.jp; $\ddagger$These authors contributed equally to this work.}
\author{Yusuke Nomura$^{2,\ddagger}$}
\author{Hiroyuki Mitamura$^{3}$}
\author{Atsushi Miyake$^{3}$}
\author{Takashi Koretsune$^{4}$}
\author{Yoshio Kaneko$^{2}$}
\author{Leonie Spitz$^{2}$}\altaffiliation{Current address: Physik-Department, Technical University of Munich, 85748 Garching, Germany}
\author{Yasujiro Taguchi$^{2}$} 
\author{Akira Matsuo$^{3}$}
\author{Koichi Kindo$^{3}$}
\author{Ryotaro Arita$^{1}$}
\author{Masashi Tokunaga$^{3}$}
\author{Yoshinori Tokura$^{1,2,5}$}
\date{\today}
\affiliation{$^{1}$Department of Applied Physics and Quantum-Phase Electronics Center (QPEC), The University of Tokyo, Bunkyo-ku, Tokyo 113-8656, Japan}
\affiliation{$^{2}$RIKEN Center for Emergent Matter Science (CEMS), Wako, Saitama 351-0198, Japan}
\affiliation{$^{3}$Institute for Solid State Physics, The University of Tokyo, 5-1-5 Kashiwanoha, Kashiwa, Chiba 277-8581, Japan}
\affiliation{$^{4}$Department of Physics, Tohoku University, Aoba-ku, Sendai, Miyagi 980-8578, Japan}
\affiliation{$^{5}$Tokyo College, The University of Tokyo, Bunkyo-ku, Tokyo 113-8656, Japan}

\begin{abstract}
When nanometric, noncoplanar spin textures with scalar spin chirality (SSC) are coupled to itinerant electrons, they endow the quasiparticle wavefunctions with a gauge field, termed Berry curvature, in a way that bears analogy to relativistic spin-orbit coupling (SOC). The resulting deflection of moving charge carriers is termed geometrical (or topological) Hall effect. Previous experimental studies modeled this signal as a real-space motion of wavepackets under the influence of a quantum-mechanical phase. In contrast, we here compare the modification of Bloch waves themselves, and of their energy dispersion, due to SOC and SSC. Using the canted pyrochlore ferromagnet Nd$_2$Mo$_2$O$_7$ as a model compound, our transport experiments and first-principle calculations show that SOC impartially mixes electronic bands with equal or opposite spin, while SSC is much more effective for opposite spin band pairs.
\end{abstract}

\maketitle

Topologically protected electronic modes are seen as an important ingredient in next-generation electronic devices for low-loss electronics and in new approaches towards quantum computation \cite{Hasan2010, Yan2017}. The existence of these is interwined with the emergent Berry curvature field $\mathbf{\Omega}_\mathbf{k}$ acting on the quantum-mechanical wave functions. For example, recent experiments on topological (semi-) metallic states in collinear ferromagnets~\cite{Ye2018, Morali2019,Belopolski2019,Liu2019,Yin2020} have established a connection between Fermi arc or drumhead surface states and excess Berry curvature in the bulk, detectable through the anomalous Hall conductivity $\sigma_{xy}^\mathrm{A}=-(e^2/h) \int \text{d}^3k \,\, \Omega_{\mathbf{k}}^z/(2\pi)^3$. It is noted that theoretical work has consistently emphasized the role of spin-orbit coupling (SOC) in magnetic topological (semi-)metals, which is believed to provide the most direct route to generating nontrivial bulk states. At the same time, calculations show that noncoplanar magnetism with nonzero scalar spin chirality (SSC) $\chi_{ijk} = \mathbf{m}_i\cdot(\mathbf{m}_j\times\mathbf{m}_k)$ of three neighboring moments $i$,$j$,$k$ can play a conceptually very similar role \cite{Ohgushi2000,Shindou2001,Martin2008}, leading to band mixing, nonzero Berry curvature, and a sideways deflection of moving charge carriers termed topological (or geometrical) Hall effect, often associated with skyrmion lattice states \cite{Bruno2004,Lee2009,Neubauer2009,Ritz2013}.

In the present work, we highlight a fundamental difference between Berry curvature generated by SOC and SSC \cite{Ohgushi2000,Shindou2001,Martin2008}. For this purpose, we separate contributions to the Berry curvature from band pairs $n\neq m$ with equal (opposite) spin as $\mathbf{\Omega}_{\mathbf{k}}^\text{diff.}=\sum_{n,m}w_{nm}^\text{diff.} \mathbf{\Omega}_{\mathbf{k}}^{nm}$ and $\mathbf{\Omega}_{\mathbf{k}}^\text{same.}=\sum_{n,m}w_{nm}^\text{same.} \mathbf{\Omega}_{\mathbf{k}}^{nm}$. Our main result is that SOC in collinear ferromagnets \cite{Fang2003, Liu2018, Chang2018, Ye2018, Gosalbez2015, Xu2018} mixes bands of equal and opposite spin, giving comparable amplitude of both $\mathbf{\Omega}_{\mathbf{k}}^\text{same}$ and $\mathbf{\Omega}_{\mathbf{k}}^\text{diff.}$ \cite{Wang2006}. In contrast, the synthetic gauge field from noncoplanar magnetic order with $\chi_{ijk}$ is related to hybridization of up and down spin states. In this scenario, excess $\mathbf{\Omega}_{\mathbf{k}}^\text{diff.}$ originates at (near-) degeneracy points for bands of opposite spin. 

The Hall signals from SOC and SSC are often referred to as Karplus-Luttinger type \cite{Karplus1954} anomalous Hall conductivity $\sigma_{xy}^\mathrm{KL}$, and as geometrical or topological Hall effect $\sigma_{xy}^\mathrm{G}$, respectively. Given a spin texture $\mathbf{m}_i$ of the conducting states, where $i$ labels discrete lattice sites, the leading terms on symmetry grounds \cite{Nagaosa2010} are $\sigma_{xy}^\mathrm{KL}\sim \sum_i m_i^z$ and $\sigma_{xy}^\mathrm{G}\sim \sum_{\langle ijk\rangle} \chi_{ijk}$. The sum $\left<ijk\right>$ is over nearest neighbor spin sites [c.f. Fig. 1(b)]. Our combined experimental and theoretical investigation provides a unified picture of Berry curvature in momentum space -- comprising both these signals. 

The target material of this study is the metallic pyrochlore Nd$_2$Mo$_2$O$_7$, a canonical canted ferromagnet \cite{Taguchi2001,Taguchi2003,Yasui2003,Yasui2006}. Its crystal structure is comprised of interpenetrating tetrahedra hosting, in turn, Mo$^{4+}$ and Nd$^{3+}$ magnetic moments. The spin arrangement on Mo-tetrahedra for two magnetic phases is illustrated in Fig. 1(a-c), demonstrating that the band picture is a reasonable starting point for this material where SSC appears with a characteristic length scale $\lambda \le a$. Here, $a$ is the lattice constant of the cubic pyrochlore structure. A large body of literature has focused on continuous spin textures with $\lambda > a$, such as skyrmions in noncentrosymmetric magnets. For such very large spin textures it is more appropriate to consider semiclassical transport under the influence of a Berry curvature field living in real space \cite{Ye1999, Bruno2004, Neubauer2009, Ritz2013}.

The ferromagnetic transition of the Mo-sublattice at $T_C  = 90\,$K is driven by the double-exchange mechanism [step-anomaly in Fig. 1(d), see Ref. \cite{Katsufuji2000}]. Meanwhile, the Nd$^{3+}$ sublattice takes a back seat at first and, upon further cooling, undergoes a smooth cross-over towards long-range order with two-in-two-out configuration on individual rare earth tetrahedra \cite{Taguchi2001, Yasui2003, Yasui2006}. This cross-over at $T^{*}\sim 20\,$K is related to rapid growth of the coercive field, visible as a departure of magnetization curves $M(T)$ recorded after cooling in $\mu_0H = 1\,$T (high-field cooled) and in $\mu_0H = 0.01\,$T (field-cooled), c.f. Fig. 1(d). The noncoplanar alignment of the Nd$^{3+}$ moments is transmitted to the conducting Mo states by moderate $d$-$f$ coupling, so that the Mo sublattice ultimately adopts canted ferromagnetic (C-FM) order with tilting angle $\sim 5-15^\circ$ away from the preferred $\left<001\right>$ easy direction [Fig. 1(a,b) and Refs. \cite{Taguchi2001,Yasui2003}]. 

Both noncoplanar and collinear arrangements of Molybdenum-spins are experimentally accessible in Nd$_2$Mo$_2$O$_7$. By use of a high magnetic field, we are able to fully align the Mo-4$d$ moments and reach a field-aligned ferromagnetic (FA-FM) state at low temperature [orange shading in Fig. 1(e,f)]. To describe the magnetization isotherms in Fig. 1(e), the net magnetization is decomposed into a sum of two contributions from Mo- and Nd-sublattices, $\mathbf{M}=\mathbf{M}_\mathrm{Nd}+\mathbf{M}_\mathrm{Mo}$. The small size of the spontaneous moment, $M_0 = 0.3\,\mathrm{\mu_B/\mathrm{f.u.}}$, is caused by near-complete cancellation between antiferromagnetically coupled $\mathbf{M}_\mathrm{Nd}$ and $\mathbf{M}_\mathrm{Mo}$ in zero field \cite{Taguchi2003}. However, the Neodymium moments can be coaligned with $\mathbf{M}_\mathrm{Mo}$ in $\mu_0H\sim 10\,$T and it is understood that the canting of Mo-spins is completely suppressed at $\mu_0H > 10\,$T \cite{Taguchi2003,Yasui2003,Yasui2006}. Note that even in FA-FM, Nd$^{3+}$ are not fully parallel to $\mathbf{H}$ up to fields $>100\,$T.

Figure 1(f), black and green curves, demonstrates that C-FM and FA-FM regimes of Nd$_2$Mo$_2$O$_7$ have distinct transport signatures, with much larger Hall conductivity $\sigma_{xy}$ in C-FM [Fig. 1(a,b)]. The Hall signal is sharply suppressed at $\mu_0H > 10\,$T when the Mo-spins are co-aligned \cite{Taguchi2003}. For further analysis of the transport properties, let us separate $\sigma_{xy}= \sigma_{xy}^\mathrm{N} + \sigma_{xy}^\mathrm{A}$, where $\sigma_{xy}^\mathrm{N}$ and $\sigma_{xy}^\mathrm{A} = \sigma_{xy}^\mathrm{KL} + \sigma_{xy}^\mathrm{G}$ are the normal and anomalous Hall conductivities, respectively. In Nd$_2$Mo$_2$O$_7$, we have $\sigma_{xy}^\mathrm{N} \propto H$ from the orbital motion of electrons and $\sigma_{xy}^\mathrm{KL} \propto M_\mathrm{Mo}$ for the SOC-driven signal. Meanwhile, $\sigma_{xy}^\mathrm{G}$ has strong field-dependence as the spin-umbrella of the C-FM order is continuously closed with increasing field [Fig. 1(a-c)]. We focus the main part of our discussion on the experimental geometry where $\mathbf{H}\,\parallelsum\,\, \left<001\right>$, but also note that the Hall conductivity shows considerable anisotropy at low magnetic fields [Fig. 1(f)]. As discussed in the Supplementary Information (SI, Ref. \cite{SI}), the low-field $\sigma_{xy}$ in the configuration $\mathbf{H}\,\parallelsum\,\left<111\right>$ is suppressed due to the effect of projecting the vector $\Omega_\mathbf{k}$ onto the $x$-$y$ plane defined by the contacts.

The heart of our study is the experimental control of the Fermi energy $\varepsilon_F$ in Nd$_2$Mo$_2$O$_7$ by chemical substitution of Ca$^{2+}$ for Nd$^{3+}$ in (Nd$_{1-x}$Ca$_x$)$_2$Mo$_2$O$_7$ (NCMO), and comparison of the observed transport response with first principle calculations. In the SI it is shown that the magnetic ground state C-FM of this compound is robust up to $x\approx 0.15$ (and $x=0.30$ for the FA-FM state), justifying our subsequent focus on the band filling effect. Fig. 1(f) includes Hall effect measurements for crystals where $\varepsilon_F$ is shifted downwards (hole doping, $x= 0.07$, violet and magenta curves). While this batch has nearly identical $T_C$ and longitudinal resistivity as compared to $x= 0$ \cite{SI}, $\sigma_{xy}(H)$ is now strongly modified; slightly weakened low-field $\sigma_{xy}^\mathrm{G}$ contrasts with significantly enhanced $\sigma_{xy}^\mathrm{KL}$. The high-field slope of $\sigma_{xy}$, i.e. $\sigma_{xy}^\mathrm{N}$, remains robust.

The anomalous Hall conductivity $\sigma_{xy}^\mathrm{A} = \sigma_{xy}^\mathrm{KL} + \sigma_{xy}^\mathrm{G}$ (after separation of $\sigma_{xy}^\mathrm{N}$) is shown in Fig. 2. As the canting angle of Mo-spins is only $\sim 10^\circ$, the dependence of $M_\mathrm{Mo}$ on magnetic field is rather weak and extrapolation from the FA-FM regime to zero field is suitable for separation of the geometrical Hall signal. We mark $\sigma_{xy}^\mathrm{KL} \sim M_\mathrm{Mo}$ by violet shading in Fig. 2. Values for the SOC- and SSC-driven Hall conductivities are thus obtained as a function of band filling $x$ [Fig. 3(a-c)]. Note that $\sigma_{xy}^\mathrm{G}$ vanishes above $x = 0.2$ in proximity to a phase transition (see SI for discussion). In contrast the FA-FM state is realized in high fields even at elevated $x$, so that Fig. 3(b) includes data of $\sigma_{xy}^\mathrm{KL}$ up to $x = 0.22$. Figure 3(a) also reports the measured low-field total Hall conductivity $\sigma_{xy}^\mathrm{0.5\,T} \approx \sigma_{xy}^\mathrm{A}$ for two directions of the magnetic field (violet and green symbols). As noted above and in SI, the low-field anisotropy arises due to a slight preference of Mo-spins for the crystallographic $\left<001\right>$ axis. 

The overall band filling dependence of $\sigma_{xy}^\mathrm{KL}$ and $\sigma_{xy}^\mathrm{G}$ can be reproduced in the framework of density functional theory. We start from the band structure in the collinear ferromagnetic state of NCMO ($x = 0$), where the size of the ordered moment is set to about $1.5\,\mu_B$/Mo~\cite{SI}, consistent with the experimental estimate \cite{Taguchi2001, Yasui2003, Taguchi2003, Yasui2006}. Electronic bands of Molybdenum 4$d$ character dominate at the Fermi edge in this calculation, and both SOC and SSC modulate the electronic structure around $\varepsilon_F$, giving rise to finite $\mathbf{\Omega}_\mathbf{k}$ and $\sigma_{xy}^\mathrm{A}$. Further changing $\varepsilon_F$ in the rigid band approximation yields the filling dependence of $\sigma_{xy}^\mathrm{KL, int.}$ and $\sigma_{xy}^\mathrm{G, int.}$ in the intrinsic regime  [Fig. 3(d,e), left-hand $y$-axes]. For comparison to the real material, the effect of quenched disorder must be taken into account using relation $\sigma_{xy}^\mathrm{KL/G} = (\sigma_{xx} / \sigma_{xx}^0)^\alpha \cdot \sigma_{xy}^\mathrm{KL/G,int.}$, with the universal exponent $\alpha= 1.6$, which is well established across a large number of material classes~\cite{SI, Onoda2008, Nagaosa2010}. Here $\sigma_{xx}^0$ is the threshold conductivity value above which the compound enters the intrinsic regime with dissipationless anomalous Hall current. While empirically $\sigma_{xx}^0 \approx 10^4\,\mathrm{\Omega^{-1}cm^{-1}}$ \cite{Onoda2008}, we tentatively use the value $0.6 \cdot 10^4\,\mathrm{\Omega^{-1}cm^{-1}}$ to get a good fit between theory and experiment for the Karplus-Luttinger term. The same correction procedure (same $\alpha$) is applied to $\sigma_{xy}^\mathrm{G,int.}$ as shown in Fig. 3(e). Further details and the possibility of slightly different $\alpha$ for $\sigma_{xy}^\mathrm{G}$ and $\sigma_{xy}^\mathrm{KL}$ are considered in SI.

Calculation and experiment show a dramatic rise of $\sigma_{xy}^\mathrm{KL}$ when increasing the hole content above $x=0$, followed by a drop when $x \ge 0.10$ [Fig. 3(b,d)]. The experimentally observed $\sigma_{xy}^\mathrm{G}$ is gently suppressed with $x$, roughly following $\sigma_{xy}^\mathrm{G}\sim (1-x)^3$. This phenomenological law \cite{Ueda2012} describes the reduction of SSC due to the dilution of rare earth moments with nonmagnetic Ca$^{2+}$. Discounting the dilution effect, we conclude that the numerical and experimental data for $\sigma_{xy}^\mathrm{G}$ both indicate very mild $x$-dependence for $x<0.15$. We stress that these calculations reproduce the sign of two components of the anomalous Hall signal, their order of magnitude, the resonance-type increase and decrease of $\sigma_{xy}^\mathrm{KL}$ as a function of $x$, and the rather mild band filling dependence of $\sigma_{xy}^\mathrm{G}$. While the former experiences a maximum around $x = 0.1$ in both experiment and theory, the experimental $\sigma_{xy}^\mathrm{KL}$ peak is narrower than that of the calculations as seen in Fig. 3(b,d). This ‘squeezing’ phenomenon can be attributed perhaps to the electron correlation effect, which is not captured in the present theory \cite{SI}. A decomposition of calculated Hall conductivities in terms of band pairs of equal spin and band pairs of opposite spin, i.e. in terms of $\mathbf{\Omega}_\mathbf{k}^\mathrm{same}$ and $\mathbf{\Omega}_\mathbf{k}^\mathrm{diff.}$, is shown in Fig. 3(d,e). While $\sigma_{xy}^\mathrm{KL}$ has roughly equal contributions from these two, $\sigma_{xy}^\mathrm{G}$ is dominated by the latter. Hence, the canted spin-structure allows strong hybridization of bands with mostly-up and mostly-down spins, but does not affect band pairs with equal spin to leading order.

We illustrate the momentum-space distribution of $\Omega_\mathbf{k}^z$ along two-dimensional cuts of the fcc cubic Brillouin zone (BZ) with $k_z = 0$ in Fig. 4(a-d). The contributions $\Omega_\mathbf{k}^\mathrm{same,z}$ and $\Omega_\mathbf{k}^\mathrm{diff.,z}$ are plotted separately and the $z$-axis is parallel to $\left<001\right>$. These contour plots, together with line cuts along the $\Gamma$-X direction in momentum space [Fig. 4(e,f)], provide an example for how SSC and SOC generate $\Omega_\mathbf{k}^z$ at different points in the BZ, and sometimes even cause opposite sign of $\Omega_\mathbf{k}^z$ in the same region of the BZ. The high sensitivity of the geometrical Hall signal $\sigma_{xy}^\mathrm{G}$ to the spin character of electronic bands is in sharp contrast to the case of the SOC-induced $\sigma_{xy}^\mathrm{KL}$. It follows that (near) half-metallic magnets with SSC, i.e. with conduction bands exclusively of spin-up type, are not well suited for the observation of enhanced $\sigma_{xy}^\mathrm{G}$. Thus, the present work gives guidance in the search for large and possibly quantized geometrical Hall responses due to SSC: The combination of complex magnetism with opposite-spin band crossings near the Fermi energy is expected to be a powerful predictor in the search for giant geometrical (or topological) Hall conductivity.

\textit{Acknowledgments.} We have benefited from discussions with N. Nagaosa, K. Ueda, I. Belopolski, and J. Masell. M.H. was supported as a Humboldt/JSPS International Research Fellow (18F18804). Y.N. acknowledges JSPS KAKENHI Grants-in-Aid for Scientific Research (No. 16H06345, No. 17K14336, and No. 18H01158). L.S. was funded by the German Academic Exchange Service (DAAD) via a PROMOS scholarship awarded by the German Federal Ministry of Education and Research (BMBF). This work was partially supported by JST CREST Grant Number JPMJCR1874 (Japan).
% Bibliography
%

\begin{figure}[!htb]
  \begin{center}
%    \includegraphics[clip, trim=9.0cm 2.5cm 9.0cm 0cm, width=1.0\linewidth]{FIG1.pdf}
%trim: [from left edge, from bottom , from right edge, ..]
		\includegraphics[clip, trim=0.5cm 0.7cm 1.2cm 1.cm, width=0.7\linewidth]{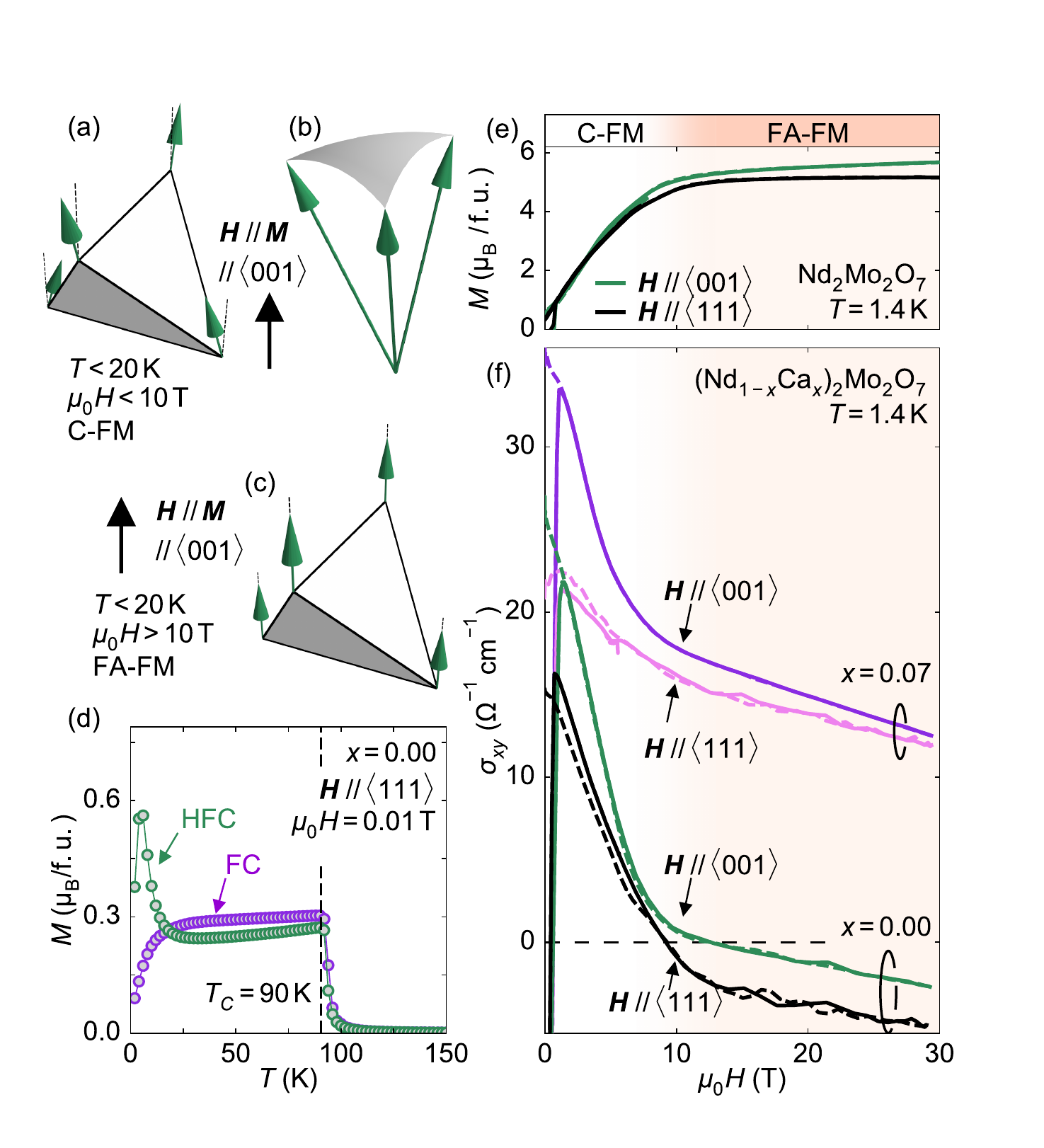}
    \caption[]{(color online). (a-c) Smooth transition of the molybdenum sublattice in (Nd$_{1-x}$Ca$_x$)$_2$Mo$_2$O$_7$ (green arrows) from (a,b) canted ferromagnet (C-FM) at low field to (c) field-aligned ferromagnet (FA-FM). (d) Magnetization $M$ as a function of increasing temperature $T$ at $x=0.00$ after preparation in a field-cooled (FC) or high-field cooled (HFC) state. Ferromagnetic ordering on the Molybdenum sublattice at $T_C$ is marked by a dashed line. (e,f) High-field magnetization and Hall conductivity measurements. The anisotropy of transport and magnetic properties is exemplified using two directions of the magnetic field $\mathbf{H}$. In (e,f), curves for increasing and decreasing magnetic field are marked by solid and dashed lines while white and orange shaded regions in (e,f) indicate the C-FM and FA-FM regimes, respectively.}
    \label{fig:fig1}
  \end{center}
\end{figure}

\begin{figure}[!b]
  \begin{center}
%    \includegraphics[clip, trim=9.0cm 2.5cm 9.0cm 0cm, width=1.0\linewidth]{FIG1.pdf}
%trim: [from left edge, from bottom , from right edge, ..]
		\includegraphics[clip, trim=0.2cm 0.6cm 1.2cm 1.cm, width=0.9\linewidth]{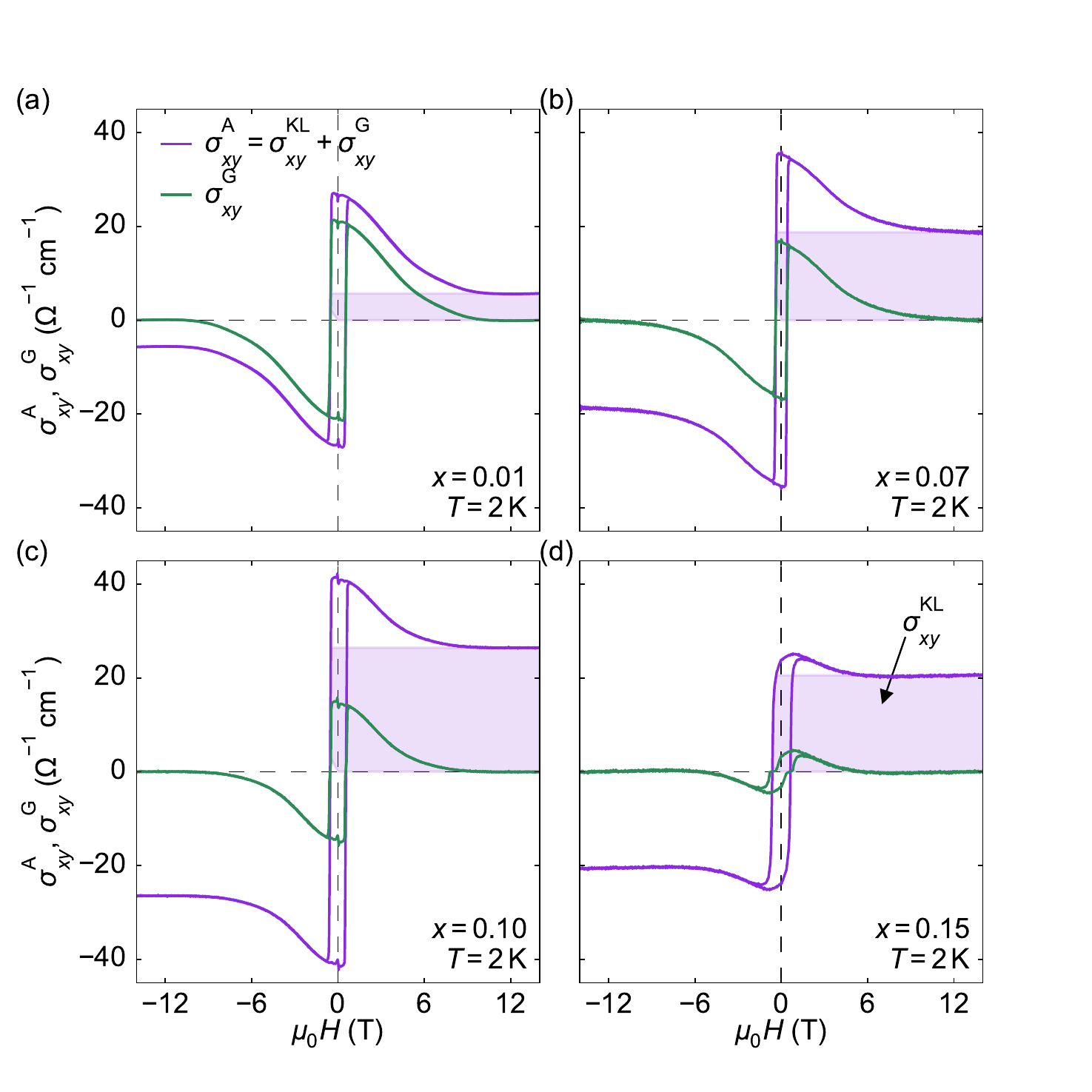}
    \caption[]{(color online). Separation of Hall conductivities in Nd$_2$Mo$_2$O$_7$ at $T = 2\,$K and $\mathbf{H} \,\parallelsum\,\, \left<001\right>$. We show the anomalous Hall conductivity $\sigma_{xy}^\mathrm{A}$ (violet curves) – after subtraction of the field-linear normal Hall signal – and the geometrical Hall signal $\sigma_{xy}^\mathrm{G}$ generated by SSC (green curves). The Hall signal originating from spin-orbit coupling, $\sigma_{xy}^\mathrm{KL}$, is marked by a violet shaded box in each plot.}
    \label{fig:fig2}
  \end{center}
\end{figure}

\begin{figure}[!htb]
  \begin{center}
%    \includegraphics[clip, trim=9.0cm 2.5cm 9.0cm 0cm, width=1.0\linewidth]{FIG1.pdf}
%trim: [from left edge, from bottom , from right edge, ..]
		\includegraphics[clip, trim=0.5cm 0.7cm 0.cm 1.6cm, width=0.9\linewidth]{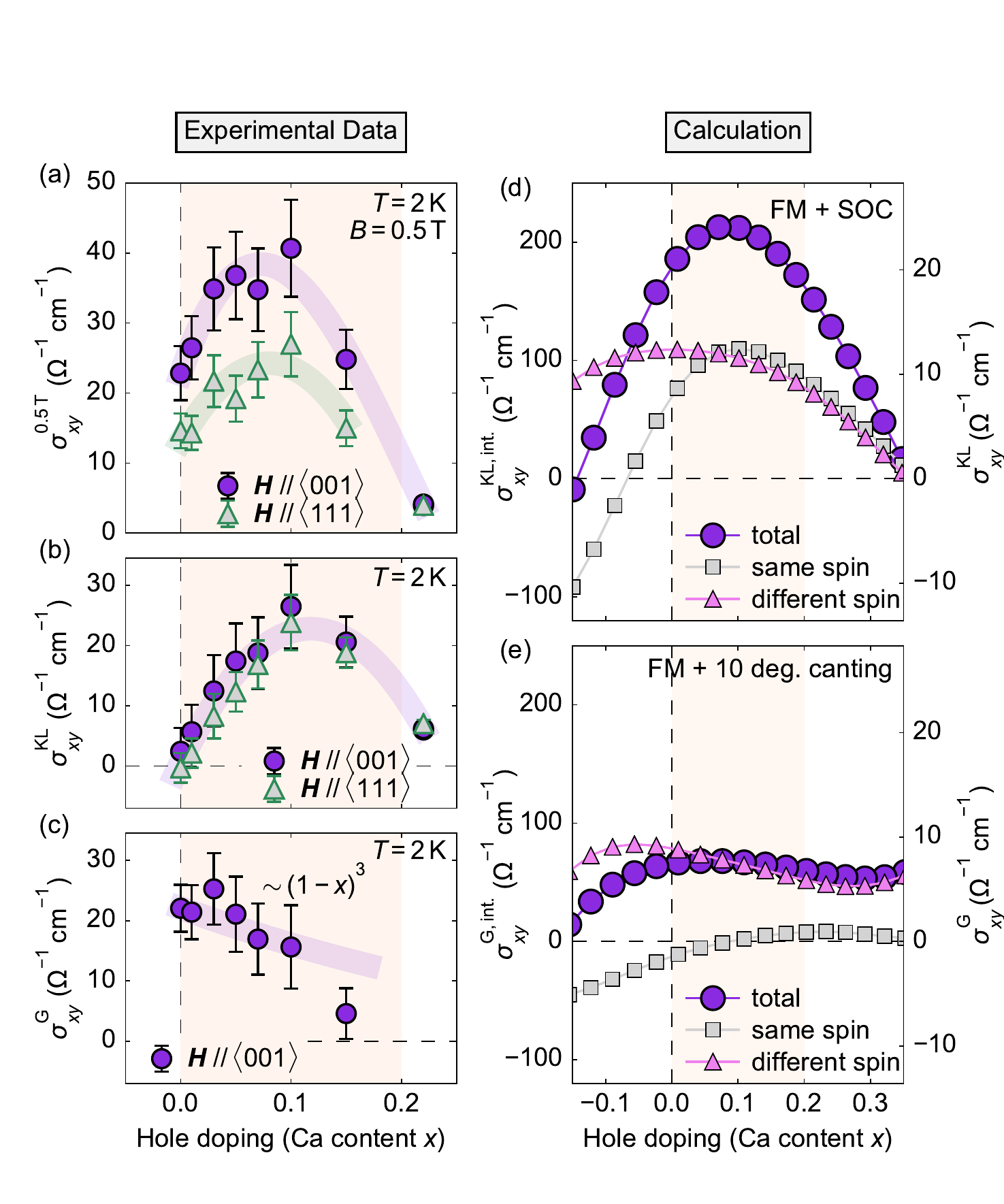}
    \caption[]{(color online). (a) Low-field Hall conductivity $\sigma_{xy}^{0.5\,\mathrm{T}}$ in (Nd$_{1-x}$Ca$_x$)$_2$Mo$_2$O$_7$, a proxy for the anomalous Hall signal $\sigma_{xy}^\mathrm{A} = \sigma_{xy}^\mathrm{G} +\sigma_{xy}^\mathrm{KL}$. (b-e) Measured and calculated SOC-driven Hall conductivity $\sigma_{xy}^\mathrm{KL}$ as well as SSC-driven geometrical Hall conductivity $\sigma_{xy}^\mathrm{G}$. In (d,e), a second $y$-axis indicates the result of the calculation ($\sigma_{xy}^\mathrm{KL,int.}$ and $\sigma_{xy}^\mathrm{G,int}$, see text) before applying the universal correction for finite defect concentration. Moreover, contributions to $\sigma_{xy}^\mathrm{KL}$, $\sigma_{xy}^\mathrm{G}$ from equal-spin and opposite-spin Berry curvature ($\mathbf{\Omega}_\mathbf{k}^\mathrm{same}$ and $\mathbf{\Omega}_\mathbf{k}^\mathrm{diff.}$) are displayed separately. Thick magenta and green lines in panels (a,b) are guides to the eye. Error bars in (a-c) are calculated from the estimated sample geometry error ($\sim \pm 17\,\% $). In (c), the thick purple line indicates the $(1-x)^3$ power law expected from dilution of rare earth magnetic moments (see text). }
    \label{fig:fig3}
  \end{center}
\end{figure}

\begin{figure*}[!htb]
  \begin{center}
%    \includegraphics[clip, trim=9.0cm 2.5cm 9.0cm 0cm, width=1.0\linewidth]{FIG1.pdf}
%trim: [from left edge, from bottom , from right edge, ..]
		\includegraphics[clip, trim=2.2cm 0.6cm 2.5cm 2.1cm, width=0.9\linewidth]{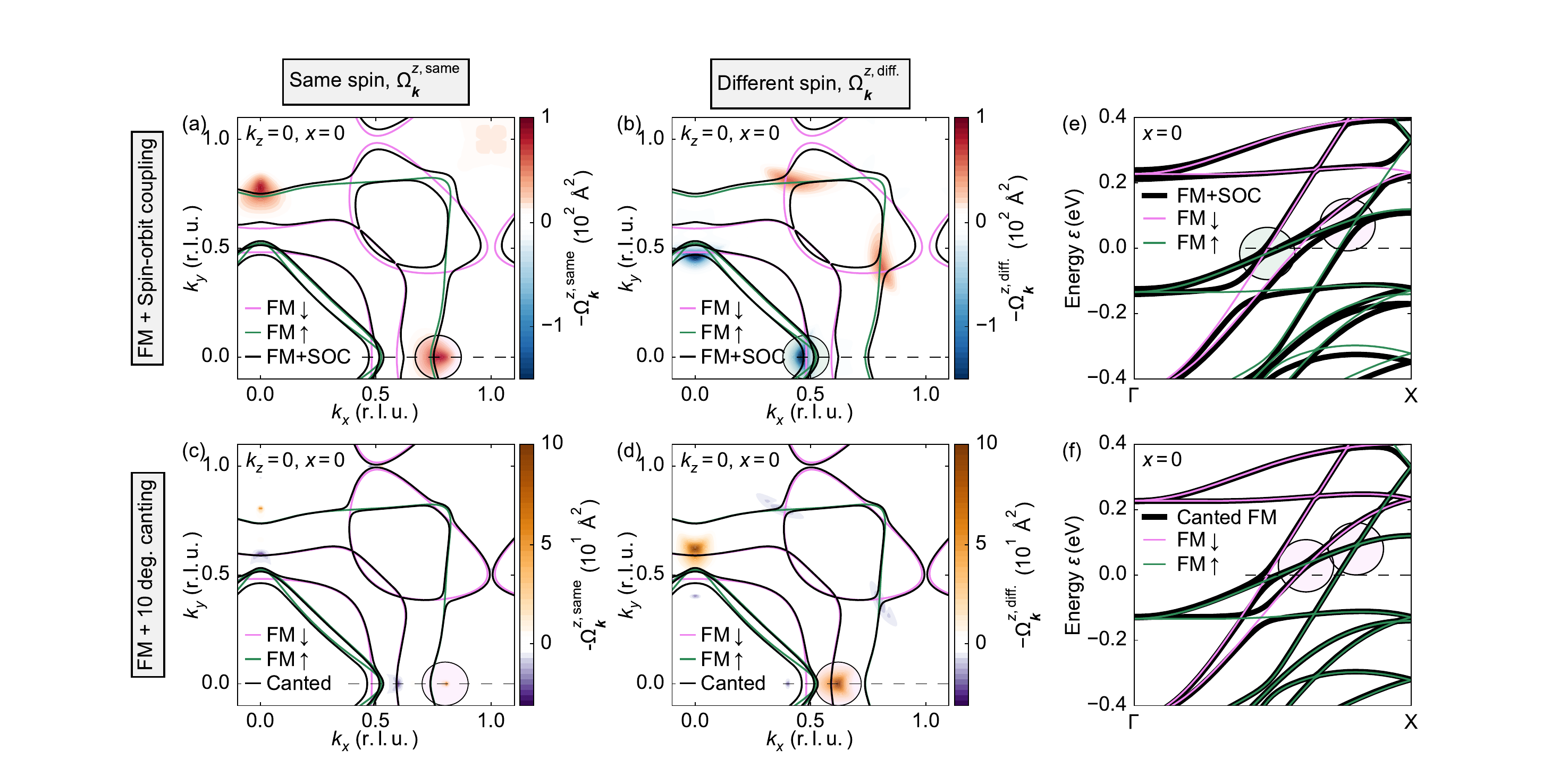}
    \caption[]{(color online). (a-d), Contour maps of Berry curvature $-\Omega_\mathbf{k}^z$ in the $k_z = 0$ plane, where same-spin and opposite-spin contributions, as defined in the text, are shown separately in two columns. The upper row is for the collinear ferromagnetic state with spin-orbit coupling (SOC), the lower row for the canted ferromagnetic (C-FM) state without SOC.  Black curves indicate Fermi surface cross-sections. For comparison, cross-sectional lines of the minority and majority Fermi surfaces in the collinear state, without SOC, are also shown. The direction of the $\Gamma$-X line in reciprocal space is indicated by a dashed line. (e,f) Along $\Gamma$-X, anti-crossing points of electronic bands are present both in the ferromagnetic state with SOC and in the C-FM state without SOC. In magenta and green, ferromagnetic bands (no SOC) are plotted for comparison. Some band-anticrossing points can be directly associated with features in the $-\Omega_\mathbf{k}^z$ maps (highlighted by shaded circles). Dashed lines mark the position of the Fermi energy $\varepsilon_F$.}
    \label{fig:fig4}
  \end{center}
\end{figure*}

\end{document}